\documentclass[runningheads]{llncs}
\usepackage{booktabs}
\usepackage{lineno}
\modulolinenumbers[5]
\usepackage{tabularx}
\usepackage{multirow}
\usepackage{pdflscape}
\usepackage[ruled,noend,norelsize,linesnumbered]{algorithm2e}
\usepackage{amsmath}
\usepackage{amssymb} 
\usepackage{bm}
\usepackage{mathtools}
\usepackage[table]{xcolor}
\usepackage{xspace} 
\usepackage{multirow}
\usepackage{setspace}
\usepackage{tikz}
\usepackage{color}
\usepackage{fullpage}
\usepackage[bookmarks=false,colorlinks]{hyperref}
\usepackage[capitalize]{cleveref}
\crefname{problem}{problem}{problems}
\Crefname{problem}{Problem}{Problems}
\usepackage{subcaption}
\usepackage{pgf,tikz,pgfplots}
\pgfplotsset{compat=1.15}
\usepackage{mathrsfs}
\usetikzlibrary{arrows}
\newtheorem{modl}{ILP Formulation for $k$-Flow Decomposition}

\newcommand{\paths}{\mathcal{P}}

\newcommand{\scs}{\mathcal{R}}

\newcommand{\Cplex}{\textsc{Cplex}\xspace}
\newcommand{\ST}{$\mathsf{Standard ILP}$\xspace}
\newcommand{\SP}{$\mathsf{Subpath Constraints ILP}$\xspace}
\newcommand{\IX}{$\mathsf{Inexact Flow ILP}$\xspace}
\newcommand{\TB}{$\mathsf{Toboggan}$\xspace}
\newcommand{\CS}{$\mathsf{Coaster}$\xspace}
\newcommand{\CSH}{$\mathsf{Coaster Heuristic}$\xspace}
\newcommand{\CF}{$\mathsf{Catfish}$\xspace}
\newcommand{\IFD}{$\mathsf{IFDSolver}$\xspace}
\newcommand{\SA}{\textbf{SRR020730-Salmon}\xspace}
\newcommand{\StT}{\textbf{SRR307903-StringTie}\xspace}
\newcommand{\RS}{\textbf{Reference-Sim}\xspace}

\usepackage{graphicx}
\begin{document}
%

\title{Fast, Flexible, and Exact Minimum Flow Decompositions via ILP\thanks{This work was partially funded by the European Research Council (ERC) under the European Union's Horizon 2020 research and innovation programme (grant agreement No.~851093, SAFEBIO), partially by the Academy of Finland (grants No.~322595, 328877, 308030), and partially by the US NSF (award 1759522).}}
%
%
\author{Fernando H. C. Dias\inst{1}\orcidID{0000-0002-6398-919X} \and
Lucia Williams\inst{2}\orcidID{0000-0003-3785-0247} \and
Brendan Mumey\inst{2}\orcidID{0000-0001-7151-2124} \and
Alexandru I. Tomescu\inst{1}\orcidID{0000-0002-5747-8350}}
\authorrunning{F. Dias et al.}
%
\institute{Department of Computer Science, University of Helsinki, Finland \email{\{fernando.cunhadias,alexandru.tomescu\}@helsinki.fi} \and
School of Computing, Montana State University, Bozeman, MT, USA
\email{\{luciawilliams,brendan.mumey\}@montana.edu}
}
\clearpage
\maketitle              
\begin{abstract}

Minimum flow decomposition (MFD) --- the problem of finding a minimum 
set of paths that perfectly decomposes a flow ---
is a classical problem in Computer Science, and variants of it are powerful models
in multiassembly problems in Bioinformatics (e.g.~RNA assembly). However, because this problem and its variants are NP-hard, practical multiassembly tools either use heuristics or solve simpler, polynomial-time solvable versions of the problem,
which may yield solutions that are not minimal or do not perfectly decompose the flow. Many RNA assemblers also use integer linear programming (ILP) formulations of such practical variants, having the major limitation they need to encode all the potentially exponentially many solution paths. Moreover, the only exact solver for MFD does not scale to large instances, and cannot be efficiently generalized to practical MFD variants. 

~~~~~~In this work, we provide the first practical ILP formulation for MFD (and thus the first fast and exact solver for MFD), based on encoding \emph{all} of the exponentially many solution paths using only a \emph{quadratic} number of variables. On both simulated and real flow graphs, our approach solves \emph{any} instance in under 13 seconds. We also show that our ILP formulation can be easily and efficiently adapted for many practical variants,
such as incorporating longer or paired-end reads, or minimizing flow errors. 

~~~~~~We hope that our results can remove the current tradeoff between the complexity of a multiassembly model and its tractability, and can lie at the core of future practical RNA assembly tools. Our implementations are freely available at \url{github.com/algbio/MFD-ILP}.

\end{abstract}

\newpage
\pagenumbering{arabic} 
\setcounter{page}{1}
\section{Introduction}

Flow decomposition (FD), the problem of decomposing
a network flow into a set of source-to-sink paths and associated weights that perfectly explain the flow values on the edges, is a classical and well-studied concept in Computer Science. For example, it is a standard result that any flow in a directed acyclic graph
(DAG) with $m$ edges can be decomposed into at most $m$ weighted paths (see,
e.g., \cite{ahuja1988network}).
However, finding an FD with a
\emph{minimum} number of paths (MFD) is NP-hard~\cite{VATINLEN20081390}, even on DAGs.
This result was later strengthened by~\cite{hartman2012split} who proved that
MFD is hard to approximate (i.e., there is some $\epsilon > 0$ such that
MFD cannot be approximated to within a $(1+\epsilon)$ factor, unless P=NP). More recent work has shown that the problem is FPT in the size of the
minimum decomposition~\cite{kloster2018practical}, and that it can be approximated
with an exponential factor~\cite{mumey2015parity}. It is also possible to decompose
all but a $\varepsilon$-fraction of the flow within a $O(1/\varepsilon)$ factor of the optimal
number of paths~\cite{hartman2012split}.
Heuristic approaches to the problem have also been developed, particularly
greedy methods based on choosing the widest or longest paths~\cite{VATINLEN20081390},
which can be improved by making iterative modifications to the flow graph before
finding a greedy decomposition~\cite{shao2017theory}. But despite this history of work on
algorithms for MFD, an exact solver that is fast for instances with large optimal solutions or large
flow values remains elusive.

\begin{sloppypar}
FD is also a key step in numerous applications. For example, some network routing problems
(e.g.~\cite{hong2013achieving,cohen2014effect,hartman2012split,mumey2015parity}) and
transportation problems~(e.g.~\cite{Ohst:2015aa,Olsen:2020aa}) require FDs that are optimal
with respect to various measures. MFDs in particular are used to
reconstruct biological sequences such as RNA transcripts~\cite{pertea2015stringtie,tomescu2013novel,gatter2019ryuto,bernard2014efficient,tomescu2015explaining,williams2019rna},
and viral quasispecies~\cite{baaijens2020strain}. However, because MFD is NP-hard,
all of these tools in fact use heuristics or solve some simpler version of the problem
ignoring some information that is available from the sequencing process,
resulting in tools that may not reconstruct the correct sequence, even if no
other errors are present. More broadly, it has been noted~\cite{nagarajan2013sequence}
that the lack of exact solvers for many of the sub-problems involved in DNA sequencing
has led to heuristic and ad hoc tools with no provable guarantees on the quality of solutions.
Additionally, some authors~\cite{bernard2014efficient,canzar2016cidane}
have noted that there is a tradeoff between the complexity of the model for RNA assembly (i.e., how much of the
true possible solution space that it supports) and its tractability. But if a fast exact solver for
MFD exists, this tradeoff may not be necessary.
\end{sloppypar}

\subsection{Minimum Flow Decomposition in Multiassembly}
\label{sec:multiassembly}

The main bioinformatics motivation for this paper is \emph{multiassembly}~\cite{xing2004multiassembly}. In this problem, we
seek to reconstruct multiple genomic sequences from mixed samples using short
substrings (called \emph{reads}) generated cheaply and accurately from next-generation sequencing
technology. The two major multiassembly problems are RNA assembly
and viral quasispecies assembly, which we describe in more detail below.

One mechanism by which complex organisms create a vast array of proteins is
alternative splicing of gene sequences, where multiple
different RNA transcripts (that are then used to produce different proteins) can be
created from the same gene~\cite{stamm2005function}. In humans, over 90\% of genes are believed
to produce multiple transcripts~\cite{wang2008alternative}.
Reconstructing the specific RNA transcripts has proved essential in characterizing gene regulation
and function, and in studying development and diseases, including cancer; see,
e.g.,~\cite{kim2008analysis,shah2012clonal}.
A second multiassembly problem is the reconstruction of viral quasispecies,
for example,
the different HIV or hepatitis strains present in a single patient sequencing sample,
or the different SARS-CoV-2 strains present in a sewage water sample. Because viruses
evolve quickly, there can be many distinct strains present at one time, and this diversity
can be an important factor in the success or effect of the virus~\cite{vignuzzi2006quasispecies}.

While the biological realities underlying the different multiassembly problems may yield
some differences in how the problems can be solved, at their heart
many approaches contain
the algorithmic step of decomposing a network flow
into weighted paths.
The basic setup and approach for multiassembly is as follows. Given a sample of
unknown sequences, each with some unknown abundance
(for example, a set of RNA transcripts or virus strains), all sequences are multiplied
and then broken into fragments that can be read by next-generation sequencers to produce
millions of sequence reads ranging from hundreds of tens of thousands DNA characters in length.
Many approaches are reference-based
(e.g., \cite{tomescu2013novel,trapnell2010transcript,maretty2014bayesian,pertea2015stringtie,kovaka2019transcriptome,bernard2014efficient,li2011isolasso}
for RNA assembly and~\cite{zagordi2011shorah,topfer2013probabilistic}
for viral quasispecies assembly),
meaning that they use a previously-constructed reference genome to guide the assembly process.
These approaches construct
a graph using the sequences contained in the reads where nodes are strings, edges represent
overlaps, and weights on edges give the counts of reads that support each overlap. Because a reference
is used, these graphs are always DAGs. In the non-reference case (called \emph{de novo}), graphs may have
cycles; we address this further at the end of the paper.  If errors
are minimal, the weights on the edges should form a flow on the network, and the underlying sequences and
their abundances must be some decomposition of the flow into weighted paths.
For RNA assembly, recent work~\cite{kloster2018practical,williams2021flow} has confirmed the common
assertion (e.g., by~\cite{tomescu2013novel,shao2017accurate,kovaka2019transcriptome,mao2020refshannon,zhao2021multitrans,lin2012cliiq,mangul2012integer})
that the true transcripts and abundances should be \emph{minimum} flow decomposition.
No such study has been done for viral quasispecies assembly,
but existing tools seek minimum-sized decompositions~\cite{baaijens2020strain,westbrooks2008hcv}.
However, while the abovementioned tools seek minimum-sized flow decompositions, since MFD is
NP-hard, they in fact
compute decompositions that are not guaranteed to be minimum
(and thus may not give the correct assembly, even when no other errors are present).

\subsection{Limitations of Current ILP Solutions}

One promising direction for fast exact solvers for MFD is integer linear programming (ILP). Existing
ILP solvers like Gurobi~\cite{gurobi} and CPLEX~\cite{studio2017cplex} incorporate optimizations
that allow for fast runtimes in practice for problems that should be hard in general (see also~\cite{gusfield2019integer} for various applications of ILP in Bioinformatics). Indeed, many
existing multiassembly tools do use ILP to solve MFD as one step in their process. The basic idea
behind these existing formulations is to consider some set of source-to-sink paths through the graph and
assign each a binary variable indicating whether or not it is selected in the optimal solution,
along with constraints to fully encode the FD problem (i.e.~that the selected set of paths---with the weights derived for them by the ILP---form an FD) and to model further practical aspects of
the specific multiassembly problem. However, the number of paths in a DAG is exponential, meaning
that if the tools enumerate all paths (and thus can be guaranteed to find the true optimal solution)
they are impractical for larger instances (e.g., Toboggan~\cite{kloster2018practical}). The most
common strategy is to pre-select some set of paths, either for all instances (e.g.,
vg-flow~\cite{baaijens2020strain} and CLIIQ~\cite{lin2012cliiq}), or only when the input is large
(e.g., MultiTrans~\cite{zhao2021multitrans} and SSP~\cite{safikhani2013ssp}). But by pre-selecting
paths, these formulations may not find the optimal MFD solution for the instance.

While the conference version of this paper was in print, the recent transcript assembly method JUMPER~\cite{jumper} was brought to our attention. JUMPER appears to be, to our knowledge, the only prior method incorporating the search for paths in a DAG into an ILP. However, their solution is slightly less general, since it works only for DAGs having a Hamiltonian path. If Hamiltonicity holds, any source-to-sink path can be encoded as a subset of edges that do not pairwise overlap in the Hamiltonian path (i.e., the tail of an edge does not appear before the head of another edge in the Hamiltonian path). As such, to avoid such pairwise edge overlaps they require a number of constraints that is quadratic in the size of the graph.

\subsection{Our Contributions}
\label{sec:our-contributions}

We give a new ILP approach to the MFD problem on DAGs, and we show that it
can be used on both simulated and real RNA assembly graphs under conditions used in many reference-based multiassembly tools.
Our approach is:

\paragraph{Fast and exact:}

We show  in \cref{sec:standard} that it is not necessary to enumerate all paths in order to encode them in
an ILP. The key idea is that any path must have a conserved (unit) flow from its start to its end,
and that this concept can be encoded using only a number of variables and constraints that is linear in the
size of the graph
(rather than exponential, as is the case when the model enumerates all possible paths).
This is a standard integer programming method for expressing paths in DAGs, used for example in~\cite{taccari2016integer}. An implementation of our ILP formulation using CPLEX finds optimal flow decomposition solutions
on RNA assembly graphs (simulated and assembled from real reads) in under 13 seconds on average, over all the datasets tested. This is
several times faster than the state-of-the-art MFD solver Toboggan~\cite{kloster2018practical}, depending on the dataset. While heuristic solvers such as Catfish~\cite{shao2017theory} or CoasterHeuristic~\cite{williams2021flow} finish withing a few seconds, we show that they do not provide optimum solutions.
Another benefit of our ILP solutions is that \emph{all} optimum solutions can be reported by the ILP solver, thus potentially helping in ``identifying'' the correct RNA multiassembly solution (a practical issue acknowledged by e.g.~\cite{findingranges,Khan:2022wo}). 

\paragraph{Flexible:}

In practice, many multiassembly tools in fact solve variants of MFD. For example, many tools account for
paired-end reads by requiring that they be included in the same path.
Another common strategy is to incorporate longer reads as subpath constraints or phasing
paths~\cite{pertea2015stringtie,shao2017accurate,williams2021flow} which again must be covered by some
predicted transcript (i.e.~flow path). In \cref{subcons}, we give additional constraints
that are expressive enough to not only encode paired-end reads and subpath constraints, but also
any generic set of edges that must be covered by a single path (e.g., as when modelling the recent Smart-seq3 protocol producing RNA multi-end reads~\cite{hagemann2020single}).
Additionally, due to sequencing or read mapping errors, the weights on edges may not
be a flow (i.e.~flow conservation might not hold). One approach in this case is to consider intervals of edge weights instead, as in~\cite{safikhani2013ssp,williams2019rna}.
We give a formulation to handle this approach in
\cref{sec:inexact}. Our implementation 
solves subpath constraint instances in similar time to standard instances, while the existing exact
solver could not complete on many instances in under 60 seconds. 
Moreover, while the existing interval heuristic is fast, it finds decompositions that are far from optimum. While all these additional constraints are naturally expressed in ILP (further underlining the flexibility of our approach), the novelty here is their integration with the ILP encoding of all possible paths in the DAG from \Cref{sec:standard}.

In \cref{sec:imperfect-flow}, we give MFD formulations dealing with 
the total error over all edges. We can consider an upper bound on the total error, or
seek a minimum decomposition that also achieves the minimum error, as
studied in~\cite{tomescu2015explaining} and used in RNA assemblers such as~\cite{li2011isolasso,li2011sparse,bernard2014efficient,tomescu2013novel}.
Finally, we note that our formulation could also be used to find decompositions for any of the above variants using a fixed, or upper bounded, number of paths, which is useful if further information is available.

\section{Preliminaries}
\label{sec:preliminaries}

Given a graph $G = (V,E)$, with vertex set $V$ and edge set $E \subseteq V \times V$, we say that $s \in V$ is a \emph{source} if $s$ has no in-coming edges. Analogously, we say that $t \in V$ is a \emph{sink} if $t$ has no out-going edges. Moreover, we say that $G$ is a \emph{directed acyclic graph (DAG)} if $G$ contains no directed cycles.

\begin{definition}[Flow network]
\label{def:flownetwork}
A tuple $G=(V,E,f)$ is said to be a \emph{flow network} if $(V,E)$ is a DAG with unique source $s$ and unique sink $t$, where for every edge $(u,v) \in E$ we have an associated positive integer \emph{flow value} $f_{uv}$, satisfying \emph{conservation of flow} for every $v \in V \setminus \{s,t\}$, namely:
\begin{equation} \label{eqn:conservation_of_flow}
\sum_{(u,v) \in E} f_{uv} = \sum_{(v,w) \in E} f_{vw}.
\end{equation}

\end{definition}

\begin{figure}[h]
\centering
\begin{subfigure}[c]{0.30\textwidth}
\includegraphics[width=\textwidth]{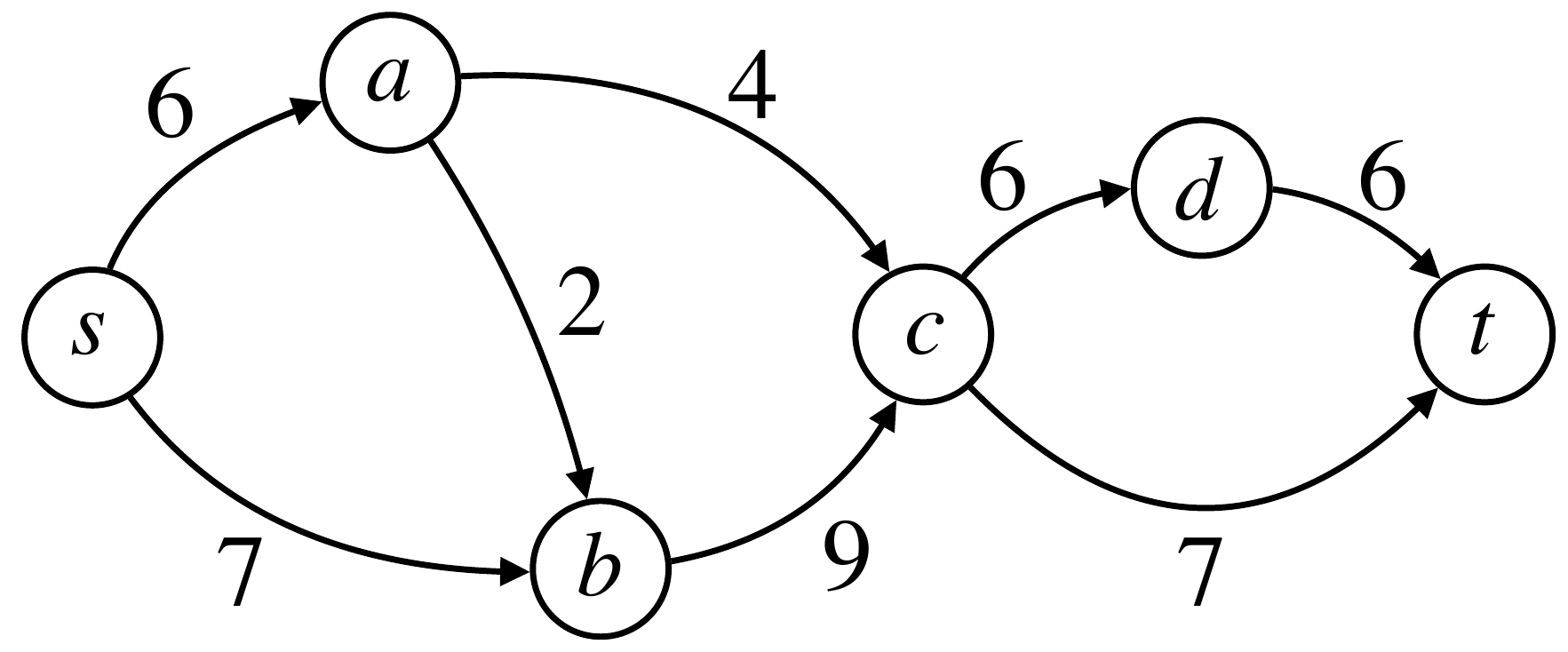}
\caption{A flow network.}
\end{subfigure}
\hfill
\begin{subfigure}[c]{0.30\textwidth}
\includegraphics[width=\textwidth]{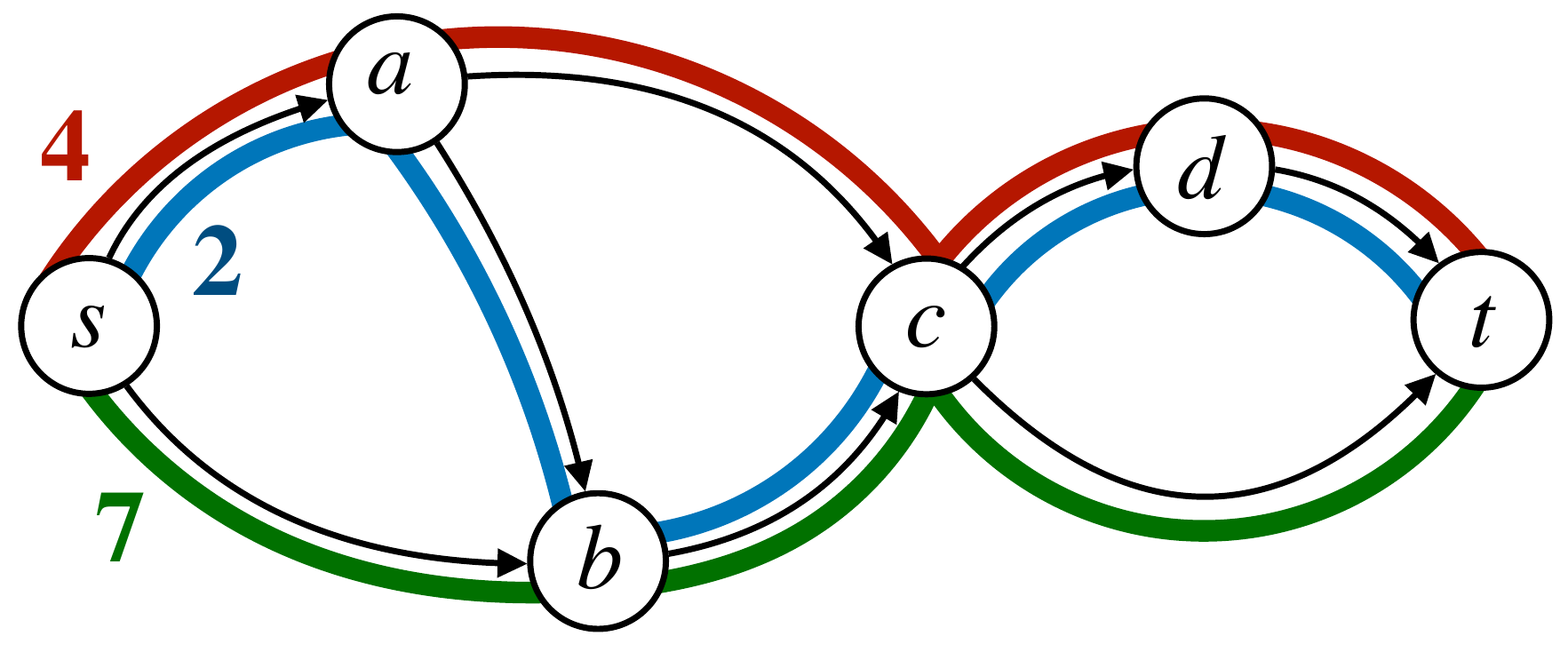}
\caption{A 3-flow decomposition into paths of weights $(4,2,7)$.\label{fig:3-FD}}
\end{subfigure}
\hfill
\begin{subfigure}[c]{0.30\textwidth}
\includegraphics[width=\textwidth]{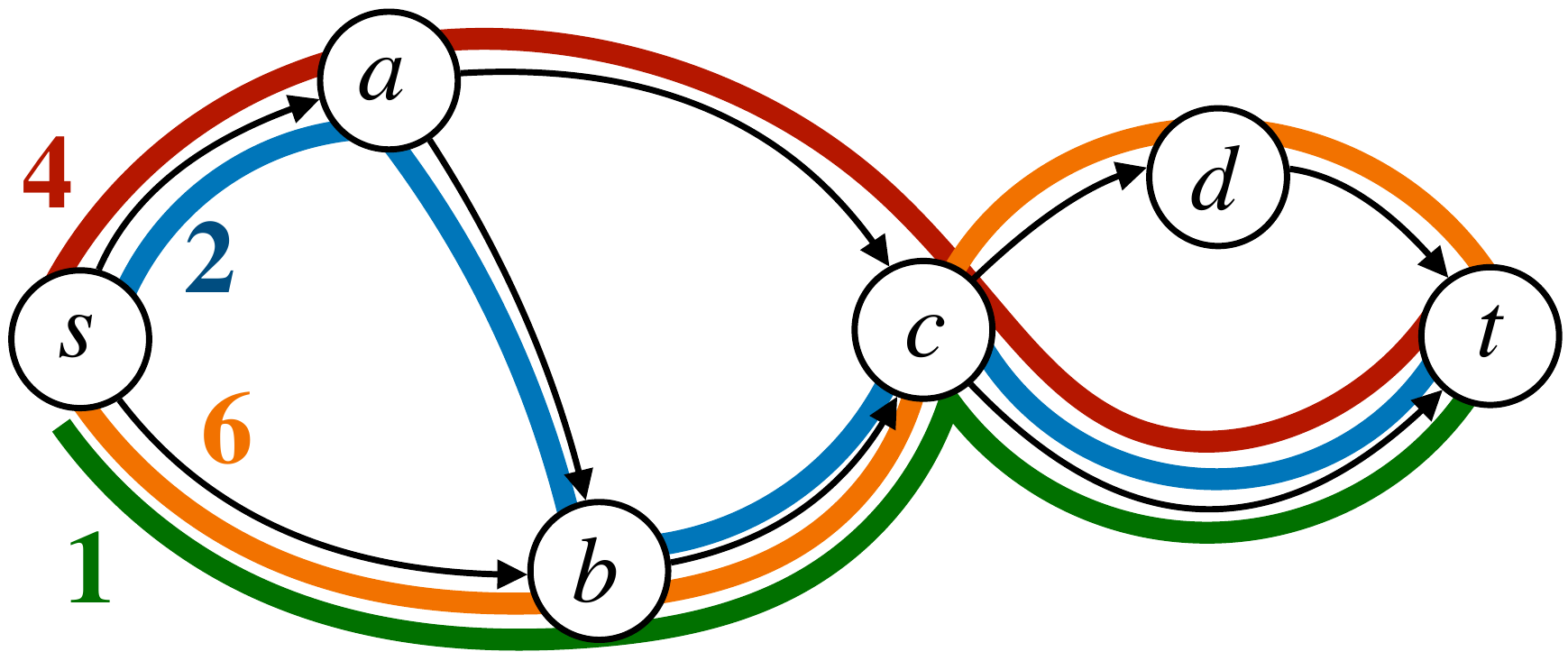}
\caption{A 4-flow decomposition into paths of weights $(4,2,6,1)$.\label{fig:4-FD}}
\end{subfigure}
\caption{Example of a flow network and of two flow decompositions of it.\label{fig:FD}}
\end{figure}

Given a flow network, a \emph{flow decomposition} for it consists of a set of source-to-sink \emph{flow paths}, and associated weights strictly greater than 0, such that the flow value of each edge equals the sum of the weights of the paths passing through that edge. In other words, the superposition of the weighted paths of the flow decomposition equals the flow of the network (see also Fig.~\ref{fig:FD}). Formally:

\begin{definition}[$k$-Flow Decomposition]
A \emph{$k$-flow decomposition} $(\paths,w)$ for a flow network $G=(V,E,f)$ is a set of $k$ $s$-$t$ flow paths $\paths = (P_1,\ldots,P_k)$
and associated weights $w = (w_1,\ldots,w_k)$,
with each $w_i \in \mathbb{Z}^{+}$, such that for each edge $(u,v) \in E$ it holds that:
\begin{equation}
\label{eqn:flow_eq}
\sum_{\substack{i \in \{1,\dots,k\} \text{ s.t. } \\ (u,v) \in P_i}} \hspace{-0.5cm}w_i = f_{uv}.
\end{equation}
\label{def:flow-decomposition}
\end{definition}

Our above definitions assume
integer flow values in the network and integer weights of the flow paths, as is natural since these values count the
number of sequenced reads traversing the edges,
and are also consistent with previous works such as~\cite{kloster2018practical}. However, in practical applications, one could have both fractional flow values and flow path weights, as in e.g.,~\cite{pertea2015stringtie}. Note also that the integer and fractional decompositions to the problem may differ. For example, \cite{VATINLEN20081390} observes that are integer flow networks which admit a $k$-flow decomposition with fractional weights, but no $k'$-flow decomposition with integer weights, for any $k' \leq k$.

\section{ILP Formulations}

\subsection{Minimum Flow Decomposition}
\label{sec:standard}

In this section we consider the following problem of finding a minimum-size flow decomposition.

\begin{problem}[Minimum flow decomposition (MFD)]
\label{prob:MFD}
Given a flow network $G=(V,E,f)$, the \emph{minimum flow decomposition (MFD) problem}
is to find a
flow decomposition $(\paths,w)$ such that $|\paths|$ is minimized.
\end{problem}

Our solution for Problem MFD is based on an ILP formulation of a flow decomposition with a given number $k$ of paths (a $k$-flow decomposition).
Using this, one can easily solve the MFD problem by finding smallest $k$ such that the flow network admits a $k$-flow decomposition. Notice that any DAG admits a flow decomposition of size at most $|E|$, see e.g.,~\cite{ahuja1988network} (since one can iteratively take the edge with smallest flow value and create an $s$-$t$ path of weight equaling this flow value).
Moreover, if assuming integer weights, another trivial upper bound on the size of any flow decomposition is $|f|$, namely the flow exiting $s$, and there is always a flow decomposition with $|f|$ paths of weight one. Thus, if there is a $k$-flow decomposition, there is also a $k'$-flow decomposition, for all $k < k' \leq \min\{|E|,|f|\}$ (just duplicate a path of weight greater than one, and move weight one from the old copy to the new one). This shows that when searching for the smaller $k$ such that the graph admits a $k$-flow decomposition we can either do a linear scan in increasing order, or binary search. Since $k$ is usually small in our applications, we just do a linear scan.
As mentioned at the end of \Cref{sec:preliminaries}, the problem can also be defined as allowing real flow values and/or weights. Our ILP formulation can also handle this variant by just changing the domain of the corresponding variables (in which case we will obtain a Mixed Integer Linear Program (MILP))\footnote{We note that this version has one subtlety to address: as discussed below, it is necessary to linearize products in the formulation to make it a true ILP (or MILP, in this case). To linearize products of the \emph{real}
variables, it is required that the real variables have closed bounds. However, if we solve $k$-FD for increasing $k$ (and not binary search), we can use $w_i \geq 0$, since no weight 0 path will be included. This introduces the limitation that this formulation could not be used to solve flow decomposition for a fixed $k$, but only if $k$ is an upper bound on the solution size.}.

We start by recalling the standard formulation of a path used for example by \cite{taccari2016integer} for the shortest path problem. If an $s$-$t$ path repeats no edge (which is always the case if the graph is a DAG) then we can interpret it simply as the set of edges belonging to the path. If we assign value 1 for each edge on the path, and value 0 for each edge not on the path, then these binary values correspond to a conceptual flow in the graph $(V,E)$ (different from the input flow). Moreover, this conceptual flow induced by the (single) path is such that the flow out-going from $s$ is 1 and the flow in-coming to $t$ is 1. It can be easily checked (cf. e.g.,~\cite{taccari2016integer}) that if the graph is a DAG, then this is a precise characterization of an $s$-$t$ path.

\begin{figure}[h]
    \centering
    \includegraphics[width=0.4\textwidth]{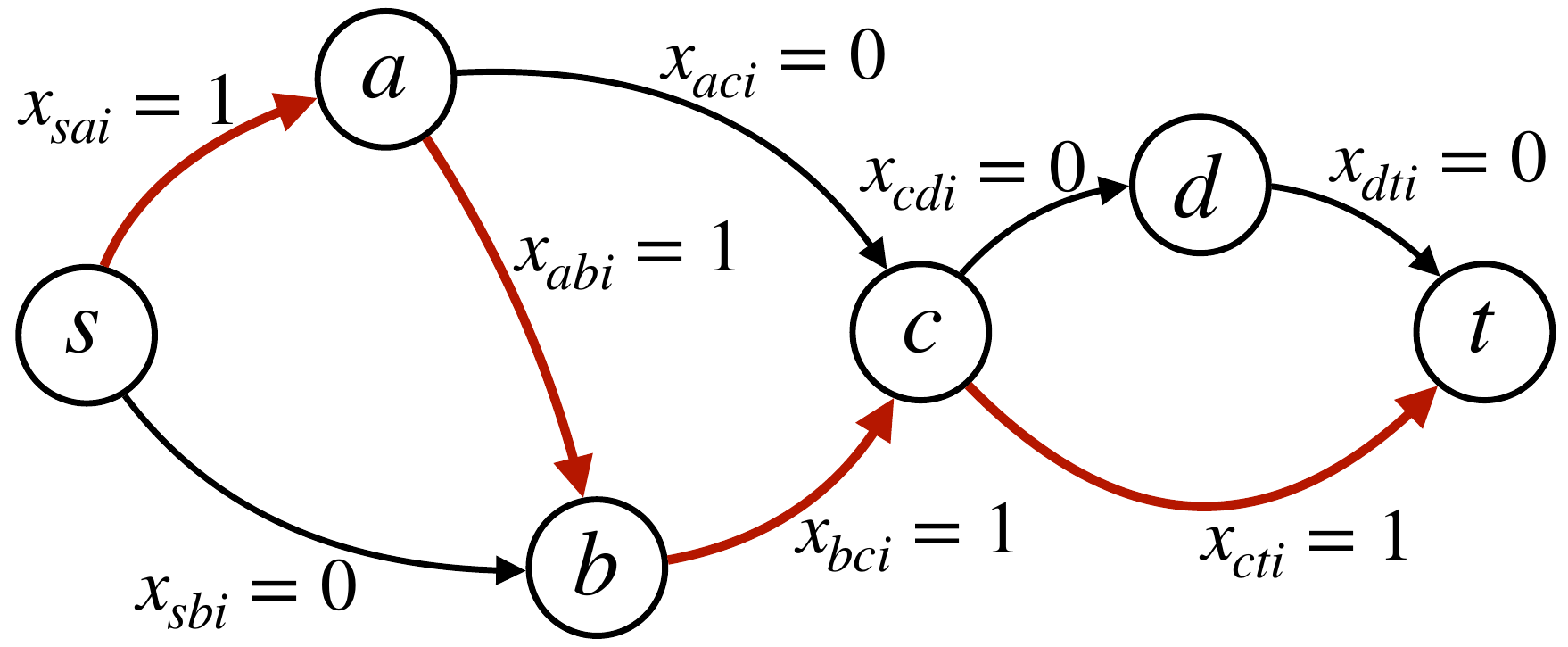}
    \caption{Example of the edge variables of the $i$\textsuperscript{th} path, satisfying \cref{s_cons,t_cons,flow_conservation_cons}.
    }
    \label{fig:FD-ILP}
\end{figure}

Thus, for every path $i \in \{1,\dots,k\}$, and every edge $(u,v) \in E$, we can introduce a binary variable $x_{uvi}$ indicating whether the edge $(u,v)$ belongs to the $i$-th path. The above characterization of a path can be expressed by the following equations (see also Fig.~\ref{fig:FD-ILP}):

\begin{subequations}
\begin{align}
   \label{s_cons}
& && \sum_{(s,v) \in E} x_{svi} = 1, &&  \forall i \in \{1, \ldots, k\}, \\
   \label{t_cons}
& && \sum_{(u,t) \in E} x_{uti} = 1, &&  \forall i \in \{1, \ldots, k\},\\ 
   \label{flow_conservation_cons}
& && \sum_{(u,v) \in E} x_{uvi} - \sum_{(v,w) \in E} x_{vwi} = 0, &&  \forall i \in \{1, \ldots, k\}, 
\forall v \in V \setminus \{s, t\}.
\end{align}
\end{subequations}

Having expressed a set of $k$ $s$-$t$ paths with already known ILP constraints, we need to introduce the new constraints tailored for the $k$-flow decomposition problem. That is, we need to state that the superposition of their weights equals the given flow in the network \eqref{eqn:flow_eq}. Thus, for each path $i$ we introduce a positive integer variable $w_i$ corresponding to its weight, and add the constraint:

\begin{align}
\label{eqn:ilp_flow_eq}
& \sum_{i \in \{1,\dots,k\}} x_{uvi}w_i = f_{uv}, && \forall (u,v) \in E.
\end{align}

To get the ILP formulation, it remains to linearize equation~\eqref{eqn:ilp_flow_eq}, which is nonlinear because it involves a product of two decision variables. Let us remark that even though non-linear programming solvers exist (such as IPOPT~\cite{wachter2006implementation}), they are inefficient, do not scale to a large number of variables and are non-professional grade. Instead, having an ILP formulation means that we can make use of popular solvers such as CPLEX \cite{studio2017cplex} and Gurobi \cite{bixby2007gurobi}.

Since the decision variables involved in the product in Eq.~\eqref{eqn:ilp_flow_eq} are bounded ($x_{uvi}$ is binary and $w_i$ is at most the largest flow value of any edge), this equation can be linearized by standard techniques as in e.g.,~\cite{furini2019theoretical} and \cite{liberti2007compact}. For that, we introduce the integer decision variable $\pi_{uvi}$ which represents the product between $w_i$ and $x_{uvi}$, and a constant $\overline{w}$ that is a large enough upper bound for any variable $w_i$ (e.g.,~the largest flow value of any edge). As such, Eq.~\eqref{eqn:ilp_flow_eq} can be replaced by the following equations: 

\begin{subequations}
\begin{align}
& && f_{uv} = \sum_{i \in \{1,\dots,k\}} \pi_{uvi}, && \forall (u,v) \in E, \label{eq:flow_lin_a}\\
& && \pi_{uvi} \leq \overline{w} x_{uvi}, && \forall (u,v) \in E, \forall i \in \{1,\dots,k\},\label{eq:flow_lin_b}\\
& && \pi_{uvi} \leq w_i, && \forall (u,v) \in E, \forall i \in \{1,\dots,k\},\label{eq:flow_lin_c}\\
& && \pi_{uvi} \geq w_i - (1-x_{uvi})\overline{w}, && \forall (u,v) \in E, \forall  i \in \{1,\dots,k\}.\label{eq:flow_lin_d}
\end{align}
\label{int_lin}
\end{subequations}

In these constraints, Eq.~\eqref{eq:flow_lin_b} ensures that $\pi_{uvi}$ is 0 if $x_{uvi}$ is 0, and Eqs.~\eqref{eq:flow_lin_c} and \eqref{eq:flow_lin_d} ensure that $\pi_{uvi}$ is $w_i$ if $x_{uvi}$ is 1. 
For completeness, see \Cref{sec:flowdecomp} for the full ILP formulation for $k$-Flow Decomposition.

\subsection{Subpath Constraints}
\label{subcons}

In this section we consider the flow decomposition variant where we are also given a set of \emph{subpath constraints} that must appear (as a subpath of some path) in any flow decomposition. Among all such decompositions we must find of one with the minimum number of paths.  In multiassembly, subpath constraints represent longer reads that span three or more vertices; they are used in popular RNA assembly tools such as StringTie~\cite{kovaka2019transcriptome} and Scallop~\cite{shao2017accurate} and their usefulness for that problem was confirmed empirically in~\cite{williams2021flow}. Such subpath constraints can also naturally model long RNA-seq reads, and we note that, as several authors also acknowledge~\cite{scallop2,amarasinghe2020opportunities,voshall2018next}, long reads do not render the RNA assembly problem obsolete, because they do not always capture full-length transcripts (due to the conversion from RNA to cDNA), and do not fully capture low-expressed transcripts.

\begin{figure}[h]
\centering
\hspace{1cm}
\begin{subfigure}[t]{0.40\textwidth}
\centering
\includegraphics[width=0.8\textwidth]{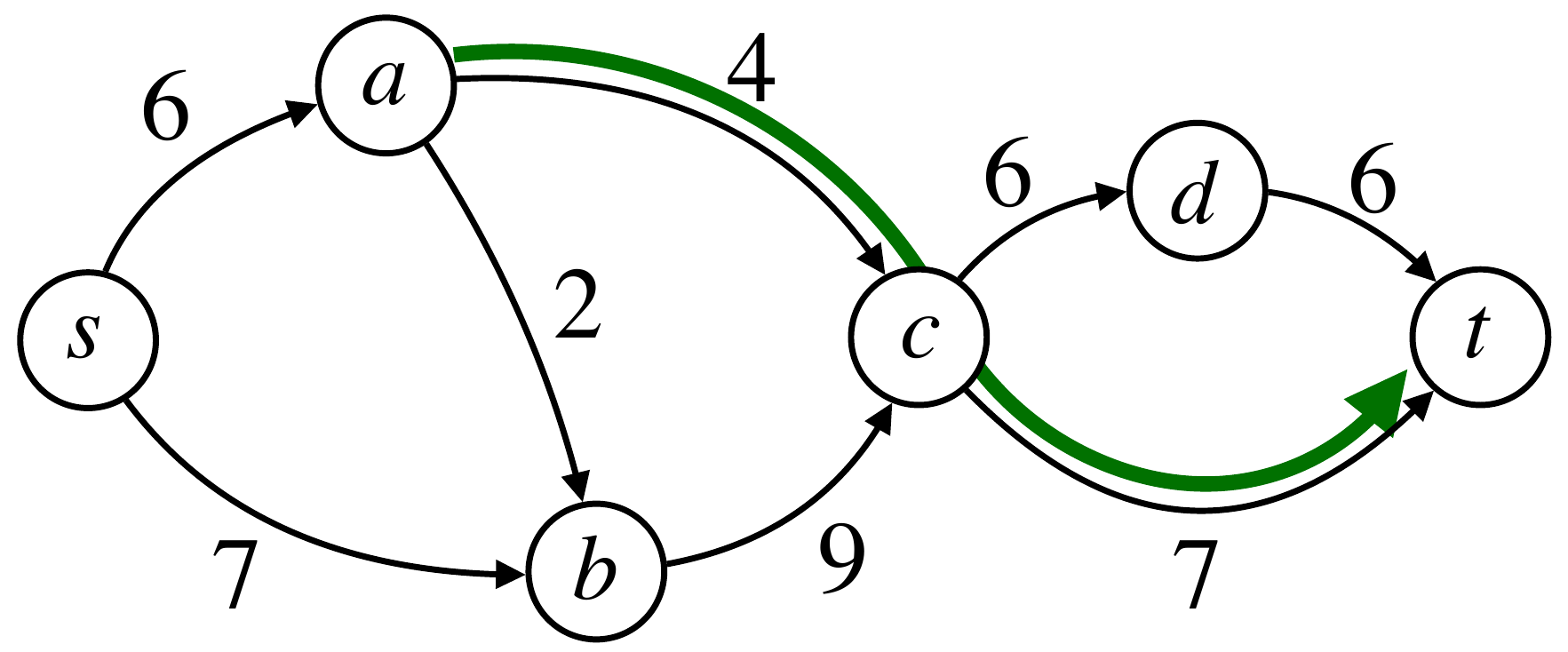}
\caption{A flow network with a single subpath constraint $R_1 = (a,c,t)$.\label{fig:FDSC-1}}
\end{subfigure}
\hfill
\begin{subfigure}[t]{0.40\textwidth}
\centering
\includegraphics[width=0.9\textwidth]{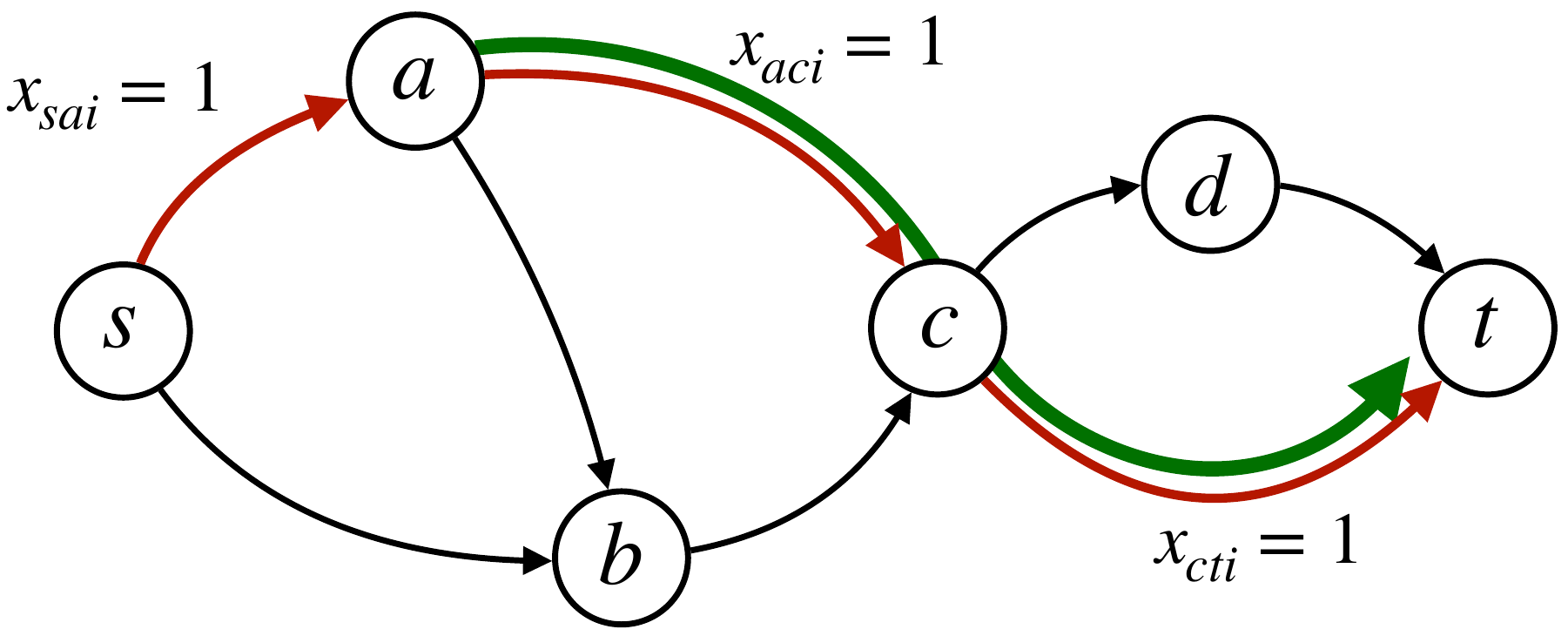}
\caption{Constraint $R_1$ is satisfied because for the $i$\textsuperscript{th} path we can set $r_{i1} = 1$ (and satisfy \cref{eq:subpath-constraint-at-least-one-path}) so that $x_{aci} + x_{cti} \geq 2r_{i1}$ holds (and satisfy \cref{eq:subpath-constraint-all-edges}).\label{fig:FDSC-2}}
\end{subfigure}
\hspace{1cm}
\caption{The flow network from Fig.~\ref{fig:FD} with a subpath constraint (which is satisfied by the 4-flow decomposition from Fig.~\ref{fig:4-FD}, but not by the one in Fig.~\ref{fig:3-FD}), and example of a path satisfying the constraint.\label{fig:FDSC}}
\end{figure}

Formally, the problem can be defined as follows (see also Fig.~\ref{fig:FDSC-1}).

\begin{definition}[Flow decomposition with subpath constraints]
Let $G=(V,E,f)$ be a flow network. \emph{Subpath constraints} are
defined to be a set of simple paths
$\scs = \{R_1,\dots,R_\ell\}$ in $G$ (not necessarily $s$-$t$ paths).
A flow decomposition $(\paths, w)$ \emph{satisfies} the subpath
constraints if and only if
\begin{equation}
\label{eqn:subpath_eq}
\forall  R_j \in \scs , \exists P_i \in \paths \mbox{ such that $R_j$ is a subpath of $P_i$.}
\end{equation}
\end{definition}

\begin{problem}[Minimum flow decomposition with subpath constraints (MFDSC)]
\label{prob:MFDSC}
Given a flow network $G=(V,E,f)$ and subpath constraints $\scs$,
the {\em minimum flow decomposition with subpath constraints} problem is to
determine if there exists, and if so, find a flow
decomposition $(\paths, w)$ satisfying $\eqref{eqn:subpath_eq}$
such that $|\paths|$ is minimized.
\end{problem}

We can expand the previous ILP formulation for $k$-Flow Decomposition to incorporate the conditions necessary to represent the subpath constraints. Let $\scs$ be the set of simple paths that are required to be part of at least one path of the flow decomposition. For each $R_j \in \scs$, we introduce an additional binary variable $r_{ij}$ denoting the presence of the subpath $R_j$ in the $i$\textsuperscript{th} path. It clearly holds that $r_{ij} = 1$ if and only if for each edge $(u,v)$ in $R_j$ we have that $x_{uvi} = 1$.
Let $|R_j|$ denote the length (i.e., number of edges) of subpath constraint $R_j$, which is a parameter (i.e.~constant). The following inequalities guarantee that each subpath constraint is satisfied by the flow decomposition (see also Fig.~\ref{fig:FDSC-2}):

\begin{subequations}
\begin{align}
& && \sum_{(u,v) \in R_j} x_{uvi} \geq |R_j| r_{ij}, && \forall i \in \{1,\dots,k\}, \forall R_j \in \scs,\label{eq:subpath-constraint-all-edges}\\
& && \sum_{i \in \{1,\dots,k\}}r_{ij} \geq 1, && \forall R_j \in \scs.\label{eq:subpath-constraint-at-least-one-path}
\end{align}
\label{subpath}
\end{subequations}

\begin{remark}
In the above ILP formulation we do not use the fact that the edges of subpath constraint $R_j$ are consecutive (i.e., form a path). Thus, the same formulation applies also if the constraint consists of a \emph{pair} of edge-disjoint paths that must all occur in the same transcript, modelling paired-end Illumina reads, or if it consists of a \emph{set} of edge-disjoint paths (or simply of a set of edges), modelling multi-end Smart-seq3 RNA reads~\cite{hagemann2020single}. More specifically, Eq.~\eqref{eq:subpath-constraint-all-edges} simply characterizes when all edges of constraint $R_j$ are covered by some flow path $i$, and Eq.~\eqref{eq:subpath-constraint-at-least-one-path} requires that at least one flow path satisfies the constraint $R_j$.
\end{remark}

\begin{remark}
While for MFD we could modify the ILP to allow also real positive path weights by setting their lower bound to be 0 (because we solve MFD by increasing $k$, as discussed at the beginning of \Cref{sec:standard}), this is no longer possible here, since the resulting model could allow as feasible \emph{optimum} solution a set of $k$ paths decomposing the flow, plus one 0-weight path added just to satisfy some subpath constraints.
\end{remark}

\subsection{Inexact Flow}
\label{sec:inexact}
Another variant of the flow decomposition problem is when the given values on the edges of the flow network do not satisfy the conservation of flow property. Instead, they are required to belong to a given interval, for each edge.
Thus, we are looking for an \emph{inexact flow decomposition}, namely one such that the superposition of its weights belongs to the given interval of each edge. This model was studied in~\cite{williams2019rna} and is used in the practical RNA assembler SSP~\cite{safikhani2013ssp}, which seeks a set of transcripts explaining the read coverage
within some user-defined error tolerance (i.e., interval around the observed weights) on all edges.

The problem is formally stated as follows.

\begin{definition}[Inexact flow network]
\label{def:inexactflownetwork}
A tuple $G=(V,E,\underline{f},\overline{f})$ is said to be an \emph{inexact flow network} if $(V,E)$ is a DAG with unique source $s$ and unique sink $t$, where for every edge $(u,v) \in E$ we have associated two positive integer values $\underline{f_{uv}}$ and $\overline{f_{uv}}$, satisfying $\underline{f_{uv}} \leq \overline{f_{uv}}$.
\end{definition}

\begin{problem}[Minimum inexact flow decomposition (MIFD)~\cite{williams2019rna}]
\label{prob:MIFD}
Given an inexact flow network $G=(V,E,\underline{f},\overline{f})$
the {\em minimum inexact flow decomposition} problem is to
determine if there exists, and if so, find 
a minimum-size set of $s$-$t$ paths $\paths = (P_1,\ldots,P_k)$
and associated weights $w = (w_1,\ldots,w_k)$
with $w_i \in \mathbb{Z}^{+}$ such that for each edge $(u,v) \in E$ it holds that:
\begin{equation}
\label{eqn:flow_eq_inexact}
\underline{f_{uv}} \leq \sum_{\substack{i \in \{1,\dots,k\} \text{ s.t. } \\ (u,v) \in P_i}} \hspace{-0.5cm}w_i \leq \overline{f_{uv}}.
\end{equation}
\end{problem}

In this variant, the same formulation as presented $k$-Flow Decomposition can be expanded to accommodate the inexact flow component. By simply replacing the flow conservation expressed in Eq.~\eqref{eqn:ilp_flow_eq} (in the linearized form in Eq.~\eqref{eq:flow_lin_a}), with the following two constraints:

\begin{subequations}
\begin{align}
   \label{eqn:inexact_flow_ilp}
& && \underline{f_{uv}} \leq  \sum_{i \in \{1,\dots,k\}} \pi_{uvi} \leq \overline{f_{uv}}, && \forall (u,v) \in E.
\end{align}
\end{subequations}

\begin{remark}
Notice that Eq.~\eqref{eqn:inexact_flow_ilp} can be combined with Eqs.~\eqref{eq:subpath-constraint-all-edges} and \eqref{eq:subpath-constraint-at-least-one-path} to obtain a solution if one needs to solve an inexact flow decomposition with subpath constraints problem, further underscoring the versatility of the ILP solution in handling various practical variants of the flow decomposition problem.
\end{remark}

\section{Experiments}


\subsection{Experiment Design}
\label{sec:design}

\paragraph{Solvers.} We denote by \ST, \SP, and \IX our ILP formulations for Problems~\ref{prob:MFD} (MFD), \ref{prob:MFDSC} (MFDSC) and \ref{prob:MIFD} (MIFD), respectively. 
We implemented these using the \Cplex Python API under default settings.
We compare \ST with \TB, the implementation by \cite{kloster2018practical} for their exact FPT algorithm for MFD, and with \CF, the implementation by \cite{shao2017theory} of their heuristic algorithm for MFD. We compare \SP with \CS, the implementation by \cite{williams2021flow} for MFDSC, which is an exact FPT algorithm extending \TB, and also with \CSH, which is a heuristic for MFDSC also by~\cite{williams2021flow}. We compare \IX with \IFD, which is an implementation of a heuristic algorithm for MIFD by \cite{williams2019rna}. Given the size of the datasets, we set a time limit for each graph, as also done by~\cite{kloster2018practical,williams2021flow} (we use 1 minute in all cases, except that we also include a run of \TB with a 5 minute time limit). The runtimes of our ILP implementations include the linear scan in increasing order to find the smallest $k$ for which there is a $k$-flow decomposition.

\paragraph{Datasets.} To test the performance of the solvers under a range of biologically-occurring graph topologies and flows weights, we used three human transcriptomic datasets containing a perfect
(i.e., the edge weights satisfy conservation of flow) splice graph for each gene of the human genome.
The first dataset, produced by the authors of~\cite{shao2017accurate} and also used
in a number of flow decomposition benchmarking studies~\cite{kloster2018practical,williams2021flow}, 
was built using publicly available RNA transcripts from the Sequence Read Archive with quantification
using the tool Salmon~\cite{patro2015salmon}. We use one of the larger transcriptomes\footnote{The full dataset
from~\cite{shao2017theory} is
available at \url{https://zenodo.org/record/1460998}. We use the file \path{rnaseq/sparse_quant_SRR020730.graph}.}
and call this dataset \textbf{SRR020730-Salmon}. We also produce perfect
splice graphs by running HiSat2~\cite{kim2019graph} with the provided GRCh38 reference index and 
then popular RNA assembly tool StringTie~\cite{kovaka2019transcriptome} on real RNA reads from SRR307903,
and superimposing the resulting transcripts and abundances (after rounding abundances to the nearest integer).
We call this dataset \textbf{SRR307903-StringTie}.
Finally, we create another dataset by directly simulating expression values for all reference transcripts of all genes
in the reference genome GRCh.104 
\textit{homo sapiens} by sampling weights from the lognormal distribution with mean $-4$ and variance $4$,
as in the default setting of the RNASeqReadSimulator tool~\cite{li2014rnaseqreadsimulator}. We multiply the
simulated values by 1000 and round to the nearest integer. We call this dataset \textbf{Reference-Sim}. For both the \textbf{Reference-Sim} and \textbf{SRR307903-StringTie} datasets, we use only genes on the positive strand.

For the subpath constraint experiments, we simulate four subpath constraints
in each graph as in \cite{williams2021flow}.
For four of the groundtruth paths,
we take the prefix of the path that includes three nontrivial junctions
(equivalent to three edges in the contracted graph described in \cite[Lemma 13]{kloster2018practical})
as a subpath constraint.
If a splice graph has fewer than four groundtruth paths, it is excluded from this experiment.

For the inexact flow experiments, we simulate interval flows as follows, similar to what was done
in~\cite{williams2019rna}. For each true edge flow $f_{uv}$, we independently sample a perturbed flow $f'_{uv}$
from $\mathcal{N}(f_{uv}, (\epsilon f_{uv})^2)$, the Gaussian
distribution with mean $f_{uv}$ and standard deviation $\epsilon f_{uv}$ 
For this experiment we fixed $\epsilon = 0.05$. We then create intervals as $[0.9 f'_{uv}, 1.1 f'_{uv}]$ with
values rounded to the nearest integer,
corresponding to a 10\% error tolerance from the observed values. As described in~\cite{williams2019rna},
it is possible that an inexact flow decomposition instance created in this way is infeasible; if an
infeasible instance is created, we re-create it until a feasible instance is found.

From all datasets, the trivial graphs made up of a single path (i.e.~admitting a trivial flow decomposition) are excluded.

\begin{table}[!h]
\centering
\caption{Results for Problem MFD.\label{tab:FD_standard}}
\resizebox{0.75\columnwidth}{!}{%
\begin{tabular}{l rr rrr rrr rrr rrrr}
\toprule
&&&\multicolumn{3}{c}{\ST} &\multicolumn{3}{c}{\TB(1 min)}& \multicolumn{3}{c}{\TB(5 min)}& \multicolumn{4}{c}{\CF}\\
\cmidrule(r){4-6} \cmidrule(r){7-9} \cmidrule(r){10-12} \cmidrule(r){13-16}
& min $k$ & Amount & Avg. & $\Sigma$  &Solved   & Avg. &  $\Sigma$ & Solved & Avg. &  $\Sigma$ & Solved  & Avg. & $\Sigma$ & Solved & Diff.\\
\midrule
\multirow{5}{*}{\rotatebox{90}{\shortstack{\textbf{SRR020730}\\\textbf{Salmon}}}}
&2-5	        &	34371	&	0.091	&	3127	&	100	&	0.002	&	68	&	100	& 0.002  & 68  & 100 	&	0.001	&	34	&	100	&	0.00	\\
&6-10        	&	2291    &	0.204	&	467	    &	100	&	0.023	&	52	&	100	& 0.024  & 54  & 100	&	0.031	&	71  &	100	&	0.00	\\
&11-15	        &   95	    &	4.692	&	445 	&	100	&	2.361	&	225	&	100	& 2.612  & 248 & 100	&	3.582	&	340	&	100	&	2.85	\\
&16-20	        &	16      &	5.891	&	94	    &	100	&	10.453	&	287	&	86	& 22.531 & 671 & 93     &	8.451	&	135	&	100	&	3.75	\\
&21-\emph{max}	&	7	    &	10.222	&	71	    &	100	&	16.564	&	281 &	50	& 33.221 & 643 & 78     &	11.621	&	81	&	100	&	4.56	\\
\midrule
\multirow{5}{*}{\rotatebox{90}{\shortstack{\textbf{Reference}\\\textbf{Sim}}}}
&2-5	       & 14513	&	0.089    & 1303	 &	100	 &	0.002	&	29	 &	100	& 0.003  & 43 & 100   &	0.058	&	841	    &	100	&	0.00	\\
&6-10	       & 1506	&	0.352    & 530	 &	100  &	0.124	&	186	 &	100	& 0.123  & 186 & 100  &	0.124	&	186	    &	100	&	0.00	\\
&11-15	       & 261	&	4.564    & 1191	 &	100	 &	24.132	&	4365 &	75	& 29.312 & 6575 & 92  &	1.299	&	339     &	100	&	2.79	\\
&16-20	       & 63	    &	10.332	 & 650   &	100	 &	36.344	&	1753 &	65	& 46.444 & 3759 & 83  &	10.45	&	658	    &	100	&	3.75	\\
&21-\emph{max} & 41	    &	12.833   & 526   &	100	 &	54.732	&	1553 &	51  & 57.672 & 4268 & 73  &	31.65	&	1298	&	100	&	4.56	\\
\midrule
\multirow{5}{*}{\rotatebox{90}{\shortstack{\textbf{SRR30790}\\\textbf{StringTie}}}}
&2-5	        &	7335	&	0.122	&	894	&	100	&	0.022	&	161	 &	100	& 0.022  & 162 & 100 &	0.029	&	212	&	100	&	0.00	\\
&6-10	        &	768	    &	1.051	&	807	&	100	&	1.191	&	914	 &	100	& 1.191  & 915 & 100 &	0.172	&	132	&	100	&	0.00	\\
&11-15	        &	133	    &	4.855	&	645	&	100	&	5.063	&	2535 &	71  & 10.343 & 5998 & 88  &	3.871	&	514	&	100	&	2.53	\\
&16-20	        &	55	    &	6.895	&	378	&	100	&	12.451	&	1764 &	57  & 21.561 & 5167 & 74  &	5.452	&	299	&	100	&	3.75	\\
&21-\emph{max}	&	37      &	10.512	&	388	&	100	&	20.562	&	1433 &	51  & 32.211 & 4362 & 68  &	9.651	&	357	&	100	&	4.56	\\
\bottomrule
\end{tabular}
}
\end{table}

\paragraph{Metrics.} For each dataset and each FD variant, we report \textbf{min $k$}, the number of paths in a minimum flow decomposition for each problem variant; \textbf{Amount}, namely the number of graphs having that specifc value of \textbf{min $k$}; \textbf{Avg.}, the average time (in seconds) for each instance solved within the time limit; $\boldsymbol{\Sigma}$, the total time (in seconds) required to solve all instances (this included also the running time of the instances that did not finish within the time limit); \textbf{Solved}, the percentage all instances solved within the  time limit; \textbf{Diff.}, the average difference between the number of paths obtained with a heuristic algorithm and the optimum one.





\subsection{Results}
\label{sec:results}

The results for Problem MFD are shown in \Cref{tab:FD_standard}.
For all three datasets, the average time and the total time of \TB and \CF outperform \ST for less complex genes, where the number of flow-paths is at most 10 or 15. However, as the genes becomes more complex (larger optimum flow decompositions), \ST is capable of solving all instances within an average of 10 seconds, while \TB and \CF require on average 16 and 11 seconds for the solved instances, respectively. In addition, \TB does not solve all instances even within the 5 minute time limit. Recall also that \CF is a heuristic, and thus it does not always return optimum solutions (see column \textbf{Diff.}). 

Among the different datasets, \SA has fewer complex genes and most instances are solved more easily. However for \StT (constructed from real RNA reads) and \RS datasets, there is a larger amount of complex genes and consequently fewer instances can be solved by \TB and \CF, while \ST remains efficient and scalable. In these results, although \ST does not perform as fast as on \SA, its runtime is still competitive, it can be scaled to graphs with larger $k$ without compromising its efficiency. On the other hand, \TB's runtime is exponential in the size of the optimum decomposition, which hinders its usage on larger instances. Moreover, notice that in some applications (e.g. cancer transcriptomics~\cite{huang2021long}) the graphs of interest do have a large number of RNA transcripts because of the genetic mechanism driving the disease.
Hence, in such applications the need to find a flow decomposition is even greater for large~$k$.

Lastly, one of the key steps in the \TB implementation is a reduction of the graph (to simplify nodes with in-degree \emph{or} out-degree equal to one, see~\cite{kloster2018practical}), which is a key insight behind its efficiency. However, this observation is highly tailored to the MFD problem, and cannot be easily extended to other FD variants (in fact, it is not used by real RNA assemblers).

The results for Problem MFDSC are shown in \Cref{tab:subpath}.
For all three datasets, \SP is capable of solving instances of any size within a few seconds. As an ILP formulation, the addition of the constraints corresponding to the subpath constraints do not hinder its scalability or efficiency. On the other hand, \CS is both slow on small instances, and does not solve large instances. This shows that the \TB implementation is optimized to use many properties of the standard MFD problem, that are not generalizable to variants of it of practical applicability, such as Problem MFDSC. Moreover, similarly to the \CF heuristic, \CSH does not return optimum solutions.


\begin{table}[!h]
\centering
\caption{Results for Problem MFDSC.\label{tab:subpath}}
\resizebox{0.7\columnwidth}{!}{%
\begin{tabular}{l rr rrr rrr rrrr}
\toprule
&&&\multicolumn{3}{c}{\SP} &\multicolumn{3}{c}{\CS} & \multicolumn{4}{c}{\CSH}\\
\cmidrule(r){4-6} \cmidrule(r){7-9} \cmidrule(r){10-13}
& min $k$ & Amount & Avg. & $\Sigma$  &Solved  & Avg. & $\Sigma$ & Solved & Avg. & $\Sigma$ & Solved & Diff.\\
\midrule
\multirow{5}{*}{\rotatebox{90}{\shortstack{\textbf{SRR020730}\\\textbf{Salmon}}}}
&4-10	        & 5691  &	0.192	&	1082 & 100 & 30.123 & 176823  & 85  & 0.005 & 28.5 &100 & 2.14 \\
&11-15	        & 95    &	1.475	&	139  & 100 & 45.121 & 4367    & 44  & 0.014 & 1.33 &100 & 3.04\\
&16-20	        & 16	& 	3.461	&	55   & 100 & 60.000 & 960     & 0   & 0.025 & 0.04 &100 & 3.91\\
&21-\emph{max}	& 8     &   10.452	&	83   & 100 & 60.000 & 480     & 0   & 0.067 & 0.536 &100 & 4.51\\
 \\
\midrule
\multirow{5}{*}{\rotatebox{90}{\shortstack{\textbf{Reference}\\\textbf{Sim}}}}
&4-10	        &	6512	& 0.18  & 1167&	100	& 37.132  & 243963 & 84 & 0.006 & 39.1 & 100 & 3.13\\
&11-15       	&	260	    & 1.10  & 279 &	100	& 46.211  & 12097  & 14 & 0.031 & 1.12 & 100 & 4.12\\
&16-20	        &	78	    & 2.58  & 203 &	100	& 60.000  & 4680   & 0  & 0.041 & 0.32 & 100 & 5.12\\
&21-\emph{max}	&	40	    & 11.51 & 460 &	100	& 60.000  & 3000   & 0  & 0.064 & 2.54 & 100 & 8.13\\
\\
\midrule
\multirow{4}{*}{\rotatebox{90}{\shortstack{\textbf{SRR30790}\\\textbf{StringTie}}}}
&4-10	        & 864		& 0.181 &  329 &	100	&  28.241 &  244001 & 86  & 0.006 & 5.18 & 100 & 2.98\\
&11-15       	& 104		& 1.124 &  148 &	100 &  45.142 &  4693  & 25  & 0.032 & 0.32 & 100 & 3.07\\
&16-20	        & 70		& 2.578 &  250 &	100	&  60.000 &  4200  & 0   & 0.083 & 0.58 & 100 & 4.14\\
&21-\emph{max}	& 27		& 11.51 &  391 &	100	&  60.000 &  1620 & 0   & 0.091 & 2.42 & 100 & 5.78\\
\\
\bottomrule
\end{tabular}
}
\end{table}

The results for Problem MIFD are shown in \Cref{tab:FD_inexact}.
For all three datasets, both formulations 
run on any instance in a small amount of time. In fact, \IX generally has the same running time as \ST, which further underscores the flexibility and efficiency of our formulations. However, \IFD is a heuristic solver, having a significant difference with respect to the size of a minimum decomposition even for small $k$.


\begin{table}[!h]
\centering
\caption{Results for Problem MIFD.\label{tab:FD_inexact}}
\resizebox{0.5\columnwidth}{!}{%
\begin{tabular}{l rr rrr rcrr}
\toprule
&&&\multicolumn{3}{c}{\IX} &\multicolumn{4}{c}{\IFD}\\
\cmidrule(r){4-6} \cmidrule(r){7-10} 
& min $k$ & Amount & Avg. & $\Sigma$  &Solved  & Avg. & $\Sigma$ & Solved & Diff.\\
\midrule
\multirow{5}{*}{\rotatebox{90}{\shortstack{\textbf{SRR020730}\\\textbf{Salmon}}}}
&2-5	     &	34371	&	0.087	&	2990 &	100	&	0.001	&	34	&	100	&	2.12\\
&6-10	     &	2291	&	0.131	&	300	&	100	&	0.025	&	57	&	100	&	2.41\\
&11-15	     &	95	    &	4.784	&	454	&	100	&	0.134	&	12	&	100	&	3.51\\
&16-20	     &	16	    &	5.784	&	91	&	100	&	0.618	&	10	&	100	&	4.13\\
&21-\emph{max}&	7       &   10.16	&	70	&	100	&	1.124	&	8	&	100	&	5.17\\
\midrule
\multirow{5}{*}{\rotatebox{90}{\shortstack{\textbf{Reference}\\\textbf{Sim}}}}
&2-5	    &	14513	&	0.153	&	2165  &	100	&	0.003	&	44	&	100	&	2.56\\
&6-10	    &	1506	&	0.109	&	164	&	100	&	0.052	&	78	&	100	&	2.78\\
&11-15	    &	261 	&	3.132   &	817	&	100	&	0.254	&	66	&	100	&	3.64\\
&16-20	    &	63	    &	5.791	&	364	&	100	&	0.783	&	50	&	100	&	3.34\\
&21-\emph{max}&	41	    &	11.56	&	473	&	100	&	1.341	&	55	&	100	&	3.56\\
\midrule
\multirow{5}{*}{\rotatebox{90}{\shortstack{\textbf{SRR30790}\\\textbf{StringTie}}}}
&2-5	      &	7335	&	0.104	&	762	&	100	&	0.001	&	7  &   100 &	2.34\\
&6-10	      &	768	    &	0.219	&	168	&	100	&	0.047	&	36	&	100	&	2.41\\
&11-15	      &	133 	&	2.891   &	384	&	100	&	0.345	&	45	&	100	&	3.40\\
&16-20	      &	55	    &	6.183	&	340	&	100	&	0.871	&	48	&	100	&	3.21\\
&21-\emph{max}&	37	    &	13.214	&	488	&	100	&	1.091	&	40	&	100	&	3.78\\
\bottomrule
\end{tabular}
}
\end{table}

\section{Conclusions and Future Work}

Flow decomposition is a key problem in Computer Science, with applications in various fields, including the major multiassembly problems from Bioinformatics. Despite this, the only exact solution for MFD is the FPT algorithm of \cite{kloster2018practical}, which does not scale to large values of $k$, and cannot be efficiently extended to model practical features of real data (such as long reads, or inexact flows). In fact, a large number of practical RNA assemblers use an ILP formulation at their core, thanks to their flexibility in modeling various aspects of real data. However, such formulations are based either on an impractical exhaustive enumeration of all possible $s$-$t$ paths, or on a greedy heuristic to select a smaller set of candidate $s$-$t$ paths that might be part of an optimum solution.

In this paper we show an efficient quadratic-size ILP for MFD and variants, avoiding for the first time the current limitation of (exhaustively) enumerating candidate $s$-$t$ paths. We also show that many constraints inside state-of-the-art RNA assemblers can be easily modeled on top of our basic ILP (i.e.~subpath constraints, inexact and imperfect flows). Further flexibility also comes from the fact that all our ILPs are based on modeling a specific type of flow decomposition with a \emph{given}, or \emph{upper bounded} number $k$ of paths (thus, they do not need to solve the minimum version of the problem). On both simulated and real datasets, we show that our ILP formulations finish within 13 seconds on any instance, and within a few seconds on most instances. 

On the practical side, we hope that our flexible ILP formulations can lie at the core of future reference-based RNA assemblers employing \emph{exact} solutions. Thus, the current tradeoff between the complexity of the model and its tractability might not be necessary anymore. On the theoretical side, our ILP formulation represents the first \emph{exact} solver for MFD scaling to large values of $k$, and it could be a reference when e.g.~benchmarking various other heuristic or approximation algorithms.


Given the maturity of ILP solvers and \TB's intrinsic exponential dependence on $k$, it is not surprising that an ILP for  MFD using a quadratic number of variables
performs significantly better than \TB for larger $k$ values. However, since for small $k$ values our ILP formulations are still slower, as future work it would be interesting to further devise more efficient MFD solvers (e.g., as a start, run \TB when the instance is detected as being ``small enough'').

It would also be interesting to extend our ILP formulations to flow networks with cycles.
While in this work we focus on reference-based approaches for multiassembly, de novo approaches
(e.g.,~\cite{grabherr2011trinity,schulz2012oases} for RNA assembly
and~\cite{baaijens2019full,baaijens2020strain,posada2021v,chen2018novo} for viral quasispecies assembly)
may yield graphs with cycles. In this context, any flow in such a network can be decomposed into at most $|E|$ weights $s$-$t$ paths and cycles. 

\bibliographystyle{splncs04}
\bibliography{biblio.bib}

\begin{thebibliography}{10}
\providecommand{\url}[1]{\texttt{#1}}
\providecommand{\urlprefix}{URL }
\providecommand{\doi}[1]{https://doi.org/#1}

\bibitem{ahuja1988network}
Ahuja, R.K., Magnanti, T.L., Orlin, J.B.: Network flows  (1988)

\bibitem{amarasinghe2020opportunities}
Amarasinghe, S.L., Su, S., Dong, X., Zappia, L., Ritchie, M.E., Gouil, Q.:
  Opportunities and challenges in long-read sequencing data analysis. Genome
  biology  \textbf{21}(1),  1--16 (2020)

\bibitem{baaijens2019full}
Baaijens, J.A., Van~der Roest, B., K{\"o}ster, J., Stougie, L., Sch{\"o}nhuth,
  A.: Full-length de novo viral quasispecies assembly through variation graph
  construction. Bioinformatics  \textbf{35}(24),  5086--5094 (2019)

\bibitem{baaijens2020strain}
Baaijens, J.A., Stougie, L., Sch{\"o}nhuth, A.: Strain-aware assembly of
  genomes from mixed samples using flow variation graphs. In: International
  Conference on Research in Computational Molecular Biology. pp. 221--222.
  Springer (2020)

\bibitem{bernard2014efficient}
Bernard, E., Jacob, L., Mairal, J., Vert, J.P.: {Efficient RNA isoform
  identification and quantification from RNA-Seq data with network flows}.
  Bioinformatics  \textbf{30}(17),  2447--2455 (2014)

\bibitem{bixby2007gurobi}
Bixby, B.: {The Gurobi Optimizer}. Transp. Re-search Part B  \textbf{41}(2),
  159--178 (2007)

\bibitem{canzar2016cidane}
Canzar, S., Andreotti, S., Weese, D., Reinert, K., Klau, G.W.: {CIDANE:
  comprehensive isoform discovery and abundance estimation}. Genome biology
  \textbf{17}(1),  1--18 (2016)

\bibitem{chen2018novo}
Chen, J., Zhao, Y., Sun, Y.: De novo haplotype reconstruction in viral
  quasispecies using paired-end read guided path finding. Bioinformatics
  \textbf{34}(17),  2927--2935 (2018)

\bibitem{cohen2014effect}
Cohen, R., Lewin-Eytan, L., Naor, J.S., Raz, D.: {On the effect of forwarding
  table size on SDN network utilization}. In: IEEE INFOCOM 2014-IEEE conference
  on computer communications. pp. 1734--1742. IEEE (2014)

\bibitem{furini2019theoretical}
Furini, F., Traversi, E.: Theoretical and computational study of several
  linearisation techniques for binary quadratic problems. Annals of Operations
  Research  \textbf{279}(1),  387--411 (2019)

\bibitem{gatter2019ryuto}
Gatter, T., Stadler, P.F.: Ry{\=u}t{\=o}: network-flow based transcriptome
  reconstruction. BMC bioinformatics  \textbf{20}(1),  1--14 (2019)

\bibitem{grabherr2011trinity}
Grabherr, M.G., Haas, B.J., Yassour, M., Levin, J.Z., Thompson, D.A., Amit, I.,
  Adiconis, X., Fan, L., Raychowdhury, R., Zeng, Q., et~al.: {Trinity:
  reconstructing a full-length transcriptome without a genome from RNA-Seq
  data}. Nature biotechnology  \textbf{29}(7), ~644 (2011)

\bibitem{gurobi}
{Gurobi Optimization, LLC}: {Gurobi Optimizer Reference Manual} (2021),
  \url{https://www.gurobi.com}

\bibitem{gusfield2019integer}
Gusfield, D.: Integer linear programming in computational and systems biology:
  an entry-level text and course. Cambridge University Press (2019)

\bibitem{hagemann2020single}
Hagemann-Jensen, M., Ziegenhain, C., Chen, P., Ramsk{\"o}ld, D., Hendriks,
  G.J., Larsson, A.J., Faridani, O.R., Sandberg, R.: {Single-cell RNA counting
  at allele and isoform resolution using Smart-seq3}. Nature Biotechnology
  \textbf{38}(6),  708--714 (2020)

\bibitem{hartman2012split}
Hartman, T., Hassidim, A., Kaplan, H., Raz, D., Segalov, M.: How to split a
  flow? In: 2012 Proceedings IEEE INFOCOM. pp. 828--836. IEEE (2012)

\bibitem{hong2013achieving}
Hong, C.Y., Kandula, S., Mahajan, R., Zhang, M., Gill, V., Nanduri, M.,
  Wattenhofer, R.: Achieving high utilization with software-driven wan. In:
  Proceedings of the ACM SIGCOMM 2013 conference on SIGCOMM. pp. 15--26 (2013)

\bibitem{huang2021long}
Huang, K.K., Huang, J., Wu, J.K.L., Lee, M., Tay, S.T., Kumar, V.,
  Ramnarayanan, K., Padmanabhan, N., Xu, C., Tan, A.L.K., et~al.: Long-read
  transcriptome sequencing reveals abundant promoter diversity in distinct
  molecular subtypes of gastric cancer. Genome Biology  \textbf{22}(1),  1--24
  (2021)

\bibitem{Khan:2022wo}
Khan, S., Kortelainen, M., C\'aceres, M., Williams, L., Tomescu, A.I.: {Safety
  and Completeness in Flow Decompositions for RNA Assembly}. CoRR
  \textbf{abs/2201.10372} (2022), \url{https://arxiv.org/abs/2201.10372}, {To}
  appear in RECOMB 2022 - Proceedings of the 26th Annual International
  Conference on Research in Computational Molecular Biology

\bibitem{kim2019graph}
Kim, D., Paggi, J.M., Park, C., Bennett, C., Salzberg, S.L.: {Graph-based
  genome alignment and genotyping with HISAT2 and HISAT-genotype}. Nature
  biotechnology  \textbf{37}(8),  907--915 (2019)

\bibitem{kim2008analysis}
Kim, P.M., Lam, H.Y., Urban, A.E., Korbel, J.O., Affourtit, J., Grubert, F.,
  Chen, X., Weissman, S., Snyder, M., Gerstein, M.B.: Analysis of copy number
  variants and segmental duplications in the human genome: Evidence for a
  change in the process of formation in recent evolutionary history. Genome
  research  \textbf{18}(12),  1865--1874 (2008)

\bibitem{kloster2018practical}
Kloster, K., Kuinke, P., O'Brien, M.P., Reidl, F., Villaamil, F.S., Sullivan,
  B.D., van~der Poel, A.: A practical fpt algorithm for flow decomposition and
  transcript assembly. In: 2018 Proceedings of the Twentieth Workshop on
  Algorithm Engineering and Experiments (ALENEX). pp. 75--86. SIAM (2018)

\bibitem{kovaka2019transcriptome}
Kovaka, S., Zimin, A.V., Pertea, G.M., Razaghi, R., Salzberg, S.L., Pertea, M.:
  {Transcriptome assembly from long-read RNA-seq alignments with StringTie2}.
  Genome biology  \textbf{20}(1),  1--13 (2019)

\bibitem{li2011sparse}
Li, J.J., Jiang, C.R., Brown, J.B., Huang, H., Bickel, P.J.: {Sparse linear
  modeling of next-generation mRNA sequencing (RNA-Seq) data for isoform
  discovery and abundance estimation}. Proceedings of the National Academy of
  Sciences  \textbf{108}(50),  19867--19872 (2011)

\bibitem{li2014rnaseqreadsimulator}
Li, W.: {RNASeqReadSimulator: a simple RNA-seq read simulator} (2014)

\bibitem{li2011isolasso}
Li, W., Feng, J., Jiang, T.: {IsoLasso: a LASSO regression approach to RNA-Seq
  based transcriptome assembly}. Journal of Computational Biology
  \textbf{18}(11),  1693--1707 (2011)

\bibitem{liberti2007compact}
Liberti, L.: Compact linearization for binary quadratic problems. 4OR
  \textbf{5}(3),  231--245 (2007)

\bibitem{lin2012cliiq}
Lin, Y.Y., Dao, P., Hach, F., Bakhshi, M., Mo, F., Lapuk, A., Collins, C.,
  Sahinalp, S.C.: Cliiq: Accurate comparative detection and quantification of
  expressed isoforms in a population. In: International Workshop on Algorithms
  in Bioinformatics. pp. 178--189. Springer (2012)

\bibitem{findingranges}
Ma, C., Zheng, H., Kingsford, C.: Finding ranges of optimal transcript
  expression quantification in cases of non-identifiability. bioRxiv  (2020).
  \doi{10.1101/2019.12.13.875625}, to appear at RECOMB 2021

\bibitem{mangul2012integer}
Mangul, S., Caciula, A., Al~Seesi, S., Brinza, D., Banday, A.R., Kanadia, R.:
  {An integer programming approach to novel transcript reconstruction from
  paired-end RNA-Seq reads}. In: Proceedings of the ACM Conference on
  Bioinformatics, Computational Biology and Biomedicine. pp. 369--376 (2012)

\bibitem{mao2020refshannon}
Mao, S., Pachter, L., Tse, D., Kannan, S.: Refshannon: A genome-guided
  transcriptome assembler using sparse flow decomposition. PloS one
  \textbf{15}(6),  e0232946 (2020)

\bibitem{maretty2014bayesian}
Maretty, L., Sibbesen, J.A., Krogh, A.: Bayesian transcriptome assembly. Genome
  biology  \textbf{15}(10),  1--11 (2014)

\bibitem{mumey2015parity}
Mumey, B., Shahmohammadi, S., McManus, K., Yaw, S.: Parity balancing path flow
  decomposition and routing. In: 2015 IEEE Globecom Workshops (GC Wkshps).
  pp.~1--6. IEEE (2015)

\bibitem{nagarajan2013sequence}
Nagarajan, N., Pop, M.: Sequence assembly demystified. Nature Reviews Genetics
  \textbf{14}(3),  157--167 (2013)

\bibitem{Ohst:2015aa}
Ohst, J.P.: On the Construction of Optimal Paths from Flows and the Analysis of
  Evacuation Scenarios. Ph.D. thesis, University of Koblenz and Landau, Germany
  (2015)

\bibitem{Olsen:2020aa}
Olsen, N., Kliewer, N., Wolbeck, L.: A study on flow decomposition methods for
  scheduling of electric buses in public transport based on aggregated
  time--space network models. Central European Journal of Operations Research
  (2020). \doi{10.1007/s10100-020-00705-6},
  \url{https://doi.org/10.1007/s10100-020-00705-6}

\bibitem{patro2015salmon}
Patro, R., Duggal, G., Kingsford, C.: {Salmon: accurate, versatile and
  ultrafast quantification from RNA-seq data using lightweight-alignment}.
  BioRxiv p. 021592 (2015)

\bibitem{pertea2015stringtie}
Pertea, M., Pertea, G.M., Antonescu, C.M., Chang, T.C., Mendell, J.T.,
  Salzberg, S.L.: {StringTie enables improved reconstruction of a transcriptome
  from RNA-seq reads}. Nature biotechnology  \textbf{33}(3),  290--295 (2015)

\bibitem{posada2021v}
Posada-C{\'e}spedes, S., Seifert, D., Topolsky, I., Jablonski, K.P., Metzner,
  K.J., Beerenwinkel, N.: V-pipe: a computational pipeline for assessing viral
  genetic diversity from high-throughput data. Bioinformatics  (2021)

\bibitem{safikhani2013ssp}
Safikhani, Z., Sadeghi, M., Pezeshk, H., Eslahchi, C.: {SSP: An interval
  integer linear programming for de novo transcriptome assembly and isoform
  discovery of RNA-seq reads}. Genomics  \textbf{102}(5-6),  507--514 (2013)

\bibitem{jumper}
Sashittal, P., Zhang, C., Peng, J., El-Kebir, M.: Jumper enables discontinuous
  transcript assembly in coronaviruses. Nature Communications  \textbf{12}(1),
  ~6728 (2021). \doi{10.1038/s41467-021-26944-y},
  \url{https://doi.org/10.1038/s41467-021-26944-y}

\bibitem{schulz2012oases}
Schulz, M.H., Zerbino, D.R., Vingron, M., Birney, E.: {Oases: robust de novo
  RNA-seq assembly across the dynamic range of expression levels}.
  Bioinformatics  \textbf{28}(8),  1086--1092 (2012)

\bibitem{shah2012clonal}
Shah, S.P., Roth, A., Goya, R., Oloumi, A., Ha, G., Zhao, Y., Turashvili, G.,
  Ding, J., Tse, K., Haffari, G., et~al.: The clonal and mutational evolution
  spectrum of primary triple-negative breast cancers. Nature
  \textbf{486}(7403),  395--399 (2012)

\bibitem{shao2017accurate}
Shao, M., Kingsford, C.: Accurate assembly of transcripts through
  phase-preserving graph decomposition. Nature biotechnology  \textbf{35}(12),
  1167--1169 (2017)

\bibitem{shao2017theory}
Shao, M., Kingsford, C.: Theory and a heuristic for the minimum path flow
  decomposition problem. IEEE/ACM transactions on computational biology and
  bioinformatics  \textbf{16}(2),  658--670 (2017)

\bibitem{stamm2005function}
Stamm, S., Ben-Ari, S., Rafalska, I., Tang, Y., Zhang, Z., Toiber, D.,
  Thanaraj, T., Soreq, H.: Function of alternative splicing. Gene
  \textbf{344},  1--20 (2005)

\bibitem{studio2017cplex}
Studio, I.I.C.O.: Cplex users manual, version 12.7 (2017)

\bibitem{taccari2016integer}
Taccari, L.: Integer programming formulations for the elementary shortest path
  problem. European Journal of Operational Research  \textbf{252}(1),  122--130
  (2016)

\bibitem{tomescu2015explaining}
Tomescu, A.I., Gagie, T., Popa, A., Rizzi, R., Kuosmanen, A., M{\"a}kinen, V.:
  Explaining a weighted dag with few paths for solving genome-guided
  multi-assembly. IEEE/ACM transactions on computational biology and
  bioinformatics  \textbf{12}(6),  1345--1354 (2015)

\bibitem{tomescu2013novel}
Tomescu, A.I., Kuosmanen, A., Rizzi, R., M{\"a}kinen, V.: {A novel min-cost
  flow method for estimating transcript expression with RNA-Seq}. In: BMC
  bioinformatics. vol.~14, pp. S15:1--S15:10. Springer (2013)

\bibitem{topfer2013probabilistic}
T{\"o}pfer, A., Zagordi, O., Prabhakaran, S., Roth, V., Halperin, E.,
  Beerenwinkel, N.: Probabilistic inference of viral quasispecies subject to
  recombination. Journal of Computational Biology  \textbf{20}(2),  113--123
  (2013)

\bibitem{trapnell2010transcript}
Trapnell, C., Williams, B.A., Pertea, G., Mortazavi, A., Kwan, G., Van~Baren,
  M.J., Salzberg, S.L., Wold, B.J., Pachter, L.: {Transcript assembly and
  quantification by RNA-Seq reveals unannotated transcripts and isoform
  switching during cell differentiation}. Nature biotechnology  \textbf{28}(5),
   511--515 (2010)

\bibitem{VATINLEN20081390}
Vatinlen, B., Chauvet, F., Chr{\'e}tienne, P., Mahey, P.: Simple bounds and
  greedy algorithms for decomposing a flow into a minimal set of paths.
  European Journal of Operational Research  \textbf{185}(3),  1390--1401
  (2008). \doi{https://doi.org/10.1016/j.ejor.2006.05.043},
  \url{https://www.sciencedirect.com/science/article/pii/S0377221706006552}

\bibitem{vignuzzi2006quasispecies}
Vignuzzi, M., Stone, J.K., Arnold, J.J., Cameron, C.E., Andino, R.:
  Quasispecies diversity determines pathogenesis through cooperative
  interactions in a viral population. Nature  \textbf{439}(7074),  344--348
  (2006)

\bibitem{voshall2018next}
Voshall, A., Moriyama, E.N.: Next-generation transcriptome assembly: strategies
  and performance analysis. Bioinformatics in the era of post genomics and big
  data pp. 15--36 (2018)

\bibitem{wachter2006implementation}
W{\"a}chter, A., Biegler, L.T.: On the implementation of an interior-point
  filter line-search algorithm for large-scale nonlinear programming.
  Mathematical programming  \textbf{106}(1),  25--57 (2006)

\bibitem{wang2008alternative}
Wang, E.T., Sandberg, R., Luo, S., Khrebtukova, I., Zhang, L., Mayr, C.,
  Kingsmore, S.F., Schroth, G.P., Burge, C.B.: Alternative isoform regulation
  in human tissue transcriptomes. Nature  \textbf{456}(7221),  470--476 (2008)

\bibitem{westbrooks2008hcv}
Westbrooks, K., Astrovskaya, I., Campo, D., Khudyakov, Y., Berman, P.,
  Zelikovsky, A.: {HCV quasispecies assembly using network flows}. In:
  International Symposium on Bioinformatics Research and Applications. pp.
  159--170. Springer (2008)

\bibitem{williams2019rna}
Williams, L., Reynolds, G., Mumey, B.: {RNA Transcript Assembly Using Inexact
  Flows}. In: 2019 IEEE International Conference on Bioinformatics and
  Biomedicine (BIBM). pp. 1907--1914. IEEE (2019)

\bibitem{williams2021flow}
Williams, L., Tomescu, A., Mumey, B.M., et~al.: Flow decomposition with subpath
  constraints. In: 21st International Workshop on Algorithms in Bioinformatics
  (WABI 2021). Schloss Dagstuhl-Leibniz-Zentrum f{\"u}r Informatik (2021)

\bibitem{xing2004multiassembly}
Xing, Y., Resch, A., Lee, C.: The multiassembly problem: reconstructing
  multiple transcript isoforms from est fragment mixtures. Genome research
  \textbf{14}(3),  426--441 (2004)

\bibitem{zagordi2011shorah}
Zagordi, O., Bhattacharya, A., Eriksson, N., Beerenwinkel, N.: Shorah:
  estimating the genetic diversity of a mixed sample from next-generation
  sequencing data. BMC bioinformatics  \textbf{12}(1), ~1--5 (2011)

\bibitem{scallop2}
Zhang, Q., Shi, Q., Shao, M.: Scallop2 enables accurate assembly of
  multiple-end rna-seq data. bioRxiv  (2021). \doi{10.1101/2021.09.03.458862},
  \url{https://www.biorxiv.org/content/early/2021/09/05/2021.09.03.458862}

\bibitem{zhao2021multitrans}
Zhao, J., Feng, H., Zhu, D., Lin, Y.: Multitrans: an algorithm for path
  extraction through mixed integer linear programming for transcriptome
  assembly. IEEE/ACM Transactions on Computational Biology and Bioinformatics
  (2021)

\end{thebibliography}

\appendix

\section{Full ILP formulation for $k$-Flow Decomposition}
\label{sec:flowdecomp}

\allowdisplaybreaks
\begin{subequations}
\begin{align}
& \sum_{(s,v) \in E} x_{svi} = 1, &&  \forall i \in \{1, \ldots, k\}, \\
& \sum_{(u,t) \in E} x_{uti} = 1, &&  \forall i \in \{1, \ldots, k\},\\ 
& \sum_{(u,v) \in E} x_{uvi} - \sum_{(v,w) \in E} x_{vwi} = 0, &&  \forall i \in \{1, \ldots, k\}, \\
& f_{uv} = \sum_{i \in \{1,\dots,k\}} \pi_{uvi}, && \forall (u,v) \in E, \\
& \pi_{uvi} \leq \overline{w} x_{uvi}, && \forall (u,v) \in E, \forall i \in \{1,\dots,k\},\\
& \pi_{uvi} \leq w_i, && \forall (u,v) \in E, \forall i \in \{1,\dots,k\},\\
& \pi_{uvi} \geq w_i - (1-x_{uvi})\overline{w}, && \forall (u,v) \in E, \forall i \in \{1,\dots,k\},\\
& w_i \in \mathbb{Z}^+, && \forall i \in \{1,\dots,k\},\\
& x_{uvi} \in \{0,1\}, && \forall (u,v) \in E, \forall i \in \{1,\dots,k\},\\
& \pi_{uvi} \in \mathbb{Z}^+ \cup \{0\}, && \forall (u,v) \in E, \forall i \in \{1,\dots,k\}.
\end{align}
\end{subequations}

\begin{table}[!ht]
\caption{Notation for $k$-Flow Decomposition ILP}
\label{notations}
\resizebox{1.0\columnwidth}{!}{%
\begin{tabular}{@{}ll@{}}  
\toprule
\textit{Headers}&                                                                                                                                      \\ \midrule
$x_{uvi}$  &  binary variable corresponding to the usage of edge $(u,v) \in E$ in flow path $i \in \{1,\dots,k\}$   \\
$w_{k}$ & integer variable corresponding to the weight of flow path $i \in \{1,\dots,k\}$ \\
$\pi_{uvi}$ & integer variable corresponding to the product of the weight of flow path $i \in \{1,\dots,k\}$ and the usage of edge $(u,v) \in E$ in the same flow path \\
$\overline{w}$ & sufficiently large upper bound for any $w_i$, for all $i \in \{1,\dots,k\}$\\
\bottomrule
\end{tabular}}
\label{tab:resultsnotations}
\end{table}


\section{Imperfect flow}
\label{sec:imperfect-flow}

An alternative approach to handle a graph whose weights to not satisfy the flow conservation property flow consists in directly taking the observed read coverages, and trying to find a set of path whose superposition best explains the observed coverages under some error model, penalizing the difference between the observed coverage of an edge and the sum of the weights of the paths going through that edge. This problem has been formalized in \cite{tomescu2015explaining} and also proven NP-hard. To formalize this problem, we denote by \emph{imperfect flow network} any DAG $(V,E)$ with unique source $s$ and unique sink $t$, where for every edge we have an associated integer positive value $f_{uv}$ (not necessarily satisfying the flow conservation property).

A first formulation of such an MFD variant imposes a fixed bound on the total 
error of all of the edges.

\begin{problem}[Minimum imperfect flow decomposition (bounded error)]
\label{prob:MimFD-BE}
Given an imperfect flow network $G=(V,E,f)$, and an error bound $B \geq 0$,
find (if it exists) a minimum-sized set of $s$-$t$ paths $\paths = (P_1,\ldots,P_k)$
and associated weights $w = (w_1,\ldots,w_k)$
with $w_i \in \mathbb{Z}^{+}$ such that for each edge $(u,v) \in E$
\begin{equation}
\label{eqn:flow_eq_imperfect}
\left|f_{uv} - \hspace{-0.3cm}\sum_{\substack{i \in \{1,\dots,k\} \text{ s.t. } \\ (u,v) \in P_i}} \hspace{-0.5cm}w_i\right| \leq B.
\end{equation}
\end{problem}

Notice that Problem~\ref{prob:MimFD-BE} is a strict generalization of the MFD problem, which is obtained by taking $B = 0$. As done in \Cref{sec:inexact}, we can obtain an ILP for it by extending the ILP formulation for $k$-Flow Decomposition to express Eq.~\eqref{eqn:flow_eq_imperfect} by the following two sets of linear equations:

\begin{align*}
& && f_{uv} - \sum_{i \in \{1,\dots,k\}} \pi_{uvi} \leq B, && \forall (u,v) \in E,\\
& && f_{uv} - \sum_{i \in \{1,\dots,k\}} \pi_{uvi} \geq -B, && \forall (u,v) \in E.
\end{align*}

This model is for a fixed $k$ value, and a full solution for Problem~\ref{prob:MimFD-BE} is obtained by trying all values of $k$ in increasing order until the ILP formulation admits a solution. Notice that the same upper bound $k \leq |E|$, since any solution to Problem~\ref{prob:MimFD-BE} (i.e.~any set of weighted $s$-$t$ paths) induces a flow, which is decomposable into at most $|E|$ weighted paths.

Another formulation, defined by~\cite{tomescu2015explaining} and at the core of RNA multiassembly tools such as \cite{li2011isolasso,li2011sparse,bernard2014efficient,tomescu2013novel}, asks to minimize the total sum of squared errors with a minimum number of paths.

\begin{problem}[Minimum imperfect flow decomposition (minimum total error)~\cite{tomescu2015explaining}]
\label{prob:MimFD-TE}
Given an imperfect flow network $G=(V,E,f)$, find a set of $s$-$t$ paths $\paths = (P_1,\ldots,P_k)$
and associated weights $w = (w_1,\ldots,w_k)$,
minimizing
\begin{equation}
\label{eqn:flow_eq_imperfect_total_error}
\sum_{(u,v) \in E}\left(f_{uv} - \sum_{\substack{i \in \{1,\dots,k\} \text{ s.t. } \\ (u,v) \in P_i}} \hspace{-0.5cm}w_i\right)^2,
\end{equation}
and among all such sets of paths, find one with minimum $k$ (i.e.~with minimum cardinality).
\end{problem}

For a given number $k$ of path, Eq.~\eqref{eqn:flow_eq_imperfect_total_error} can be used as an objective function in an Integer Quadratic Problem (IQP), which can solved by commercial solvers such as CPLEX and Gurobi. The main requirements is that the objective function is quadratic and convex, such as:

\begin{align}
\label{eqn:objqp}
& \min \sum_{(u,v) \in E}\left(f_{uv} - \sum_{i \in \{1,\dots,k\}} \pi_{uvi}\right)^2.
\end{align}

As before, to fully solve Problem~\ref{prob:MimFD-TE}, one can iterate over $k$ from 1 to $|E|$ (upper bound holding by the same reasoning as above), and choose the smallest one attaining Eq.~\eqref{eqn:objqp}.

\end{document}